\documentclass[aps,prstab, amsmath, amssymb, preprint, superscriptaddress,10pt]{revtex4-1}

\usepackage{amssymb}
\usepackage{graphicx}
\usepackage{subfigure}

\usepackage{amsmath}
\usepackage{color}

\begin{document}




\title{Controlling the spectral shape of nonlinear Thomson scattering with proper laser chirping}



\author{S.G.~Rykovanov}
\email{S.Rykovanov@gsi.de}
\altaffiliation{Presently at Helmholtz-Institut Jena, Fr\"obelstieg 3, 07743, Jena, Germany}
\address{Lawrence Berkeley National Laboratory, Berkeley, CA 94720}

\author{C.G.R.~Geddes}
\address{Lawrence Berkeley National Laboratory, Berkeley, CA 94720}

\author{C.B.~Schroeder}
\address{Lawrence Berkeley National Laboratory, Berkeley, CA 94720}

\author{E.~Esarey}
\address{Lawrence Berkeley National Laboratory, Berkeley, CA 94720}

\author{W.P.~Leemans}
\address{Lawrence Berkeley National Laboratory, Berkeley, CA 94720}

\begin{abstract}
Effects of nonlinearity in Thomson scattering of a high intensity laser pulse from electrons are analyzed.  Analytic expressions for laser pulse shaping in amplitude and frequency are obtained which control spectrum broadening for arbitrarily high laser pulse intensities.   These analytic solutions allow prediction of the spectral form and required laser parameters to avoid broadening.  The predictions are validated by numerical calculations.  This control over the scattered radiation bandwidth allows  of narrow bandwidth sources to be produced using high scattering intensities, which in turn greatly improves scattering yield for future x- and gamma-ray sources.

\end{abstract}

\maketitle





\section{Introduction}
\label{intro}

Scattering of laser light from fast moving electrons (in this paper referred to as Thomson scattering or TS) is a  widely used source of X- and gamma-ray photons~\cite{Sprangle1993, Nedorezov2004, Achterhold2013,Albert2010,Sarachik1970,Arutyunyan1964,Bulyak2002,Chen1994,Esarey1993,Hartemann2007,Kawase2008,Kulikov1964,Leemans1996,Litvinenko1996, Ohgaki1994, Pogorelsky2000,Schoenlein1996, Schwoerer2006, Villa1996}. Applications include photon sources of  percent-level bandwidth for Nuclear Resonance Fluorescence or photofission (\cite{Bertozzi2008,Albert2011a,Quiter2011a,Rykovanov2013} and references therein), for radiography~\cite{Quiter2008,Tommasini2011} and for medical applications~\cite{Weeks1997c,Carroll2003}. Thomson sources' monochromaticity, wide-range tunability (depending on the energy of the electrons in the beam) and directionality provide important advantages over Bremsstrahlung sources.

Applications of Thomson sources require high photon fluxes, which is challenging due to the small Thomson cross section.   The total number of electrons that can be accelerated in a bunch is limited for the high quality bunches required to produce narrow bandwidth sources.  For example, Laser Plasma Accelerator are typically near $N_e\sim 10^8$~\cite{Esarey2009}, and  conventional linacs are typically in the same range for low emittances~\cite{Ding2009}. 
 Scattering laser performance and geometry are then  principal tools for achieving the required fluxes.  The produced photon yield is proportional to the product of $N_e$  with the scattering laser intensity and pulse length.  The need to match the laser diffraction range to its pulse length (to keep intensity constant over the scattering volume), has meant that increasing yield by increasing laser pulse length costs quadratically in laser energy.  For example, to double yield, pulse length must double and diffraction range must also double, requiring the spot size to increase by $\sqrt{2}$.   This makes it desirable to scatter at  high  laser intensity in order to maximize yield at reasonable laser energy cost.   
 
 Scattering laser intensity is strongly limited by the fact that the generated spectrum can be broadened and band-like structure can appear in the fundamental frequency as well as its harmonics~\cite{Hartemann1996, Hartemann2010, Brau2004,Seipt2011,Heinzl2010, Hartemann2013, Ghebregziabher2013}  even for rather low values of laser pulse amplitude $a_0=eA_L/m_ec^2$ on the order of $a_0>0.1$, where $e$ and $m_e$ are charge and mass of the positron respectively, $c$ is the speed of light in vacuum and $A_L$ is the laser vector potential amplitude. This is a detrimental and limiting effect in the cases when a narrow bandwidth gamma or x-ray sources is essential, especially in the case of Nuclear Resonance Fluorescence for active nuclear interrogation. It puts a limit on the maximum laser intensity and thus maximum laser pulse amplitude $a_0$ that can be used for obtaining a certain FWHM bandwidth of the photon source. For example, such \textit{moderate} laser pulse amplitude as $a_0=0.2$ already leads to broadening on the order of 4$\%$.  The limit on intensity is an important driver of the laser energy and hence cost of a Thomson source: producing one photon per electron costs 1.6 Joules of laser energy at $a_0 =0.15$, but only 0.4 J at $a_0 =0.3$.

Limited sets of parameters have been found to minimize the nonlinear effects have been suggested based on shaping of the laser pulse temporal envelope.  Hartemann \textit{et al}~\cite{Hartemann1996} 

showed that a flat top longitudinal intensity profile could reduce the effects of nonlinearity, but this approach is typically limited by the diffraction range of the laser pulse.  

In fact, FELs work using undulator strength parameters $a_0$ on the order of unity for the optimal photon yield.   Three-dimensional effects that appear due to tight laser pulse focusing can be also detrimental in TS as discussed by Hartemann~\textit{et al}~\cite{Hartemann2013}. Ideally one would need laser pulses having near flat-top distribution in all dimensions. 
 Ghebregziabher \textit{et al}~\cite{Ghebregziabher2013} have proposed controlling the shape of the photon spectrum using laser pulse chirping and demonstrated it using a single set of numerical simulations, and such simulations have been extended by Terzic \textit{et al}~\cite{Terzic2014} and the latter results were published while this manuscript was being prepared.   There is however need for an analytical framework and solutions for nonlinear broadening, spectral shape, and compensation of these effects using laser pulse shaping in amplitude and frequency to obtain narrow bandwidth.     

In this manuscript we derive analytical expressions for the spectrum of the nonlinear Thomson scattering for both the case of the unchirped and chirped pulses and compare them to the results of numerical integration. We demonstrate that proper laser pulse chirping leads to bandwidth narrowing for arbitrarily large laser pulse amplitudes.   This allows design of sources for narrow bandwidth together with high efficiency, without scanning of numerical parameters.  The paper is organized as follows. In Sec.~\ref{broadening} a physical explanation of the spectrum broadening and appearance of the sub-structures in the spectrum is provided. Analytical expressions for the nonlinear Thomson scattering spectrum are derived and compared with the results of the numerical integration. In Sec.~\ref{chirping_sec} this analysis is extended to show that the laser frequency-versus-time  dependence, or chirp, can be used to compensate for nonlinear effects and to narrow the spectrum even for high intensity lasers. Finally, discussions and conclusions are presented in Sec.~\ref{discussion_sec}.

\section{Nonlinear broadening and spectral shape } \label{broadening}
Spectral broadening and appearance of sub-structures due to the nonlinear effects in Thomson scattering has been reported in several previous works \cite{Hartemann1996, Hartemann2010, Brau2004,Seipt2011,Heinzl2010, Ghebregziabher2013,Terzic2014}.  Of note is that this process is also alternately referred to as Compton Scattering in the literature,  but as we neglect the recoil on electrons in this analysis we use the term Thomson Scattering herein.  The appearance of  substructures was identified to be due to constructive and destructive interference of radiation emitted from different electron positions within the scattering laser pulse. Here we begin by reviewing this physical explanation of the shape of the spectrum, then derive the shape of the spectrum analytically and compare it to that obtained via numerical integration.  In addition to providing analytic derivation of the spectrum, this will serve as the basis for the derivation of compensation techniques required to produce narrow bandwidth at high intensity. 

As is common practice, we work here in the frame of reference where an electron is initially at rest and the electromagnetic laser wave impinges it. Results presented can be immediately applied to the case of scattering from the electron moving with arbitrary speed (e.g.  a relativistic electron beam) by using Lorentz transform.  Similarly, we use a circularly polarized  scattering laser as it allows us to obtain analytical expressions for the spectrum similar to solutions derived by Hartemann \textit{et al}~\cite{Hartemann1996}. In terms of number of generated photons there is no difference for the case of pulses with different polarizations with same energy. The case of linear polarization can be modeled in the straightforward way using numerical integration~\cite{Heinzl2010, Ghebregziabher2013}.

\begin{figure}[h!!]
\subfigure{\includegraphics[width=0.4\textwidth]{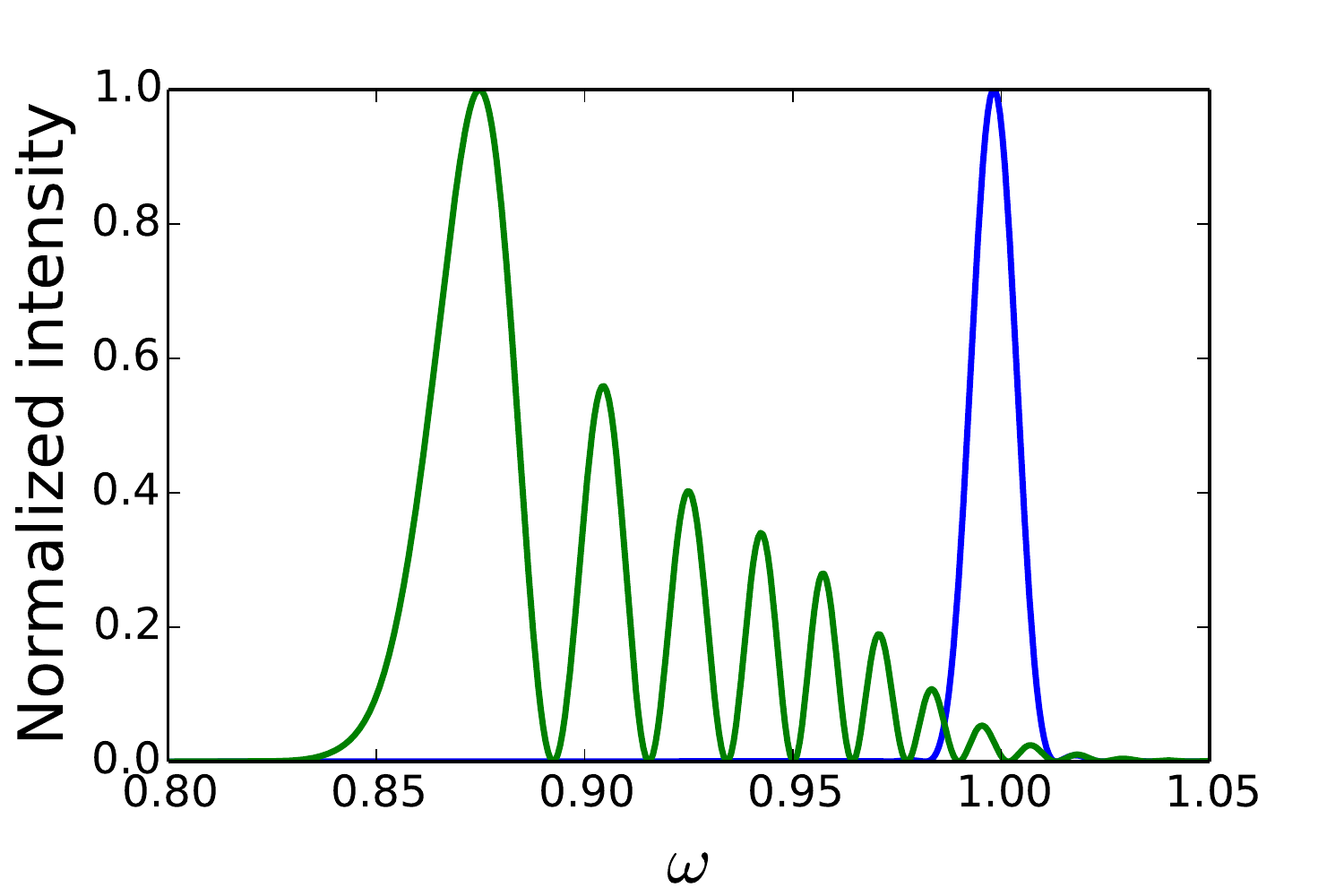}}
\subfigure{\includegraphics[width=0.4\textwidth]{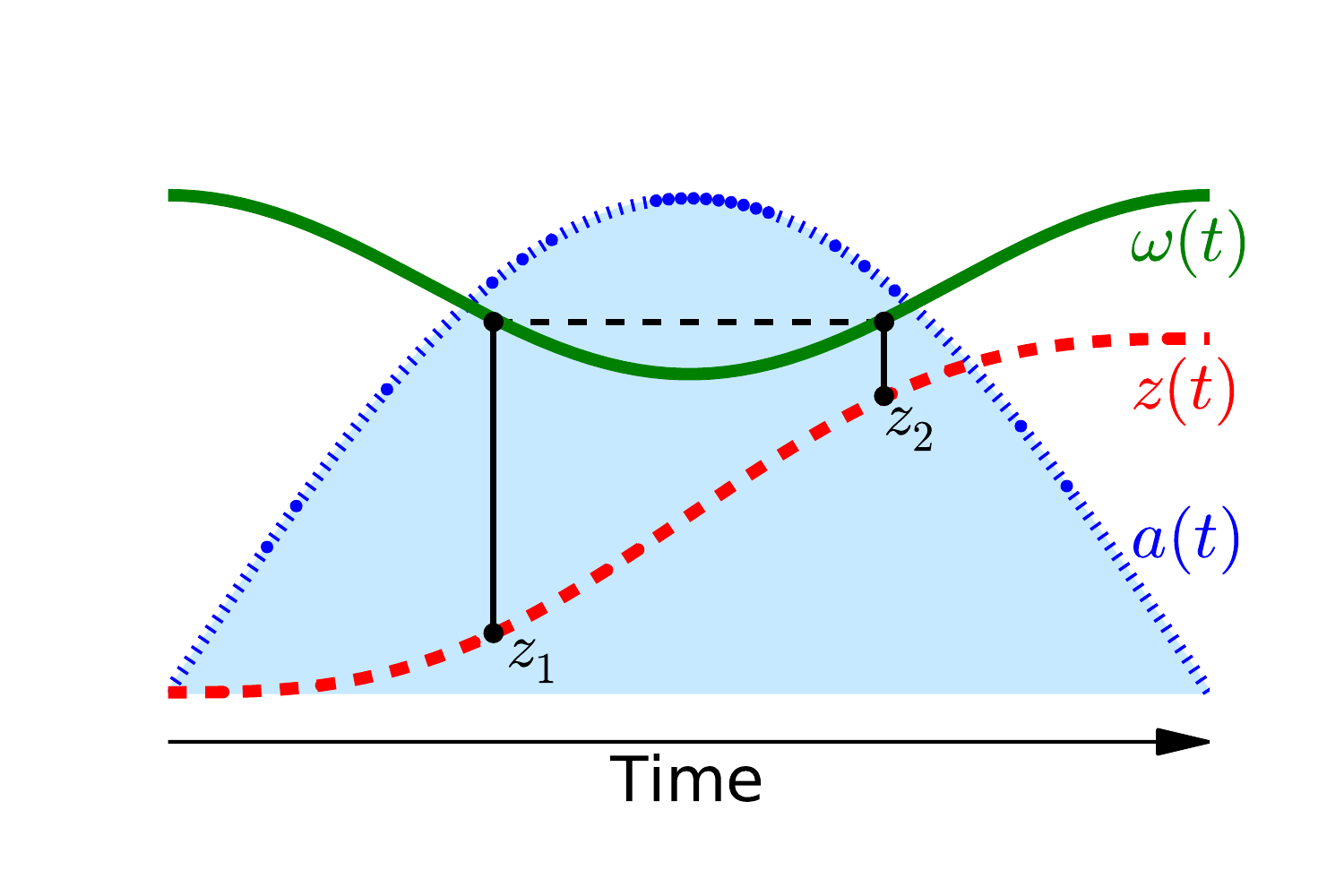}}
\caption{Left: An example of the normalized spectra of reflected radiation calculated for low $a_0=0.05$ (blue line) and high $a_0=0.4$ (red color) scattering lasers, demonstrating the appearance of band-like structure in the spectrum. Right:  Qualitative illustration of the broadening and band formation mechanism in the nonlinear response of an electron to a strong electromagnetic wave. A laser impinges the electron from the left side (from $z\rightarrow -\infty$). As functions of time, the blue line and shaded area represent the laser pulse envelope, the green line shows the frequency of the reflected wave in accordance with eq.~(\ref{generated_freq_eq}), and the red dashed line shows the longitudinal coordinate of the electron.  }
\label{schematic_figure}
\end{figure}

The spectrum produced is considerably broadened in the case of the strong laser pulse, and a band-like structure appears as shown in Fig.~\ref{schematic_figure}~(left) for $a_0=0.4$, as compared with the linear case $a_0\ll 1$ (in this case $a_0=0.05$).   This can be understood qualitatively from the equation describing the frequency of the photon reflected from an electron which for the case of on-axis scattering and assuming circular laser polarization is given by
\begin{equation}
\omega =\frac{1}{1+a_0^2}\mathrm{,}\label{generated_freq_eq}
\end{equation}
where $\omega=\omega'/\omega_L$ with $\omega'$ and $\omega_L$ being the generated and laser frequencies respectively both in CGS units.  The convention used in this paper is that circular polarized laser pulse  has twice the energy of the linearly polarized laser pulse with same $a_0$, so that in the case of linear polarization the frequency is given by $\omega=\frac{1}{1+a_0^2/2}$.  Note also that eq.~(\ref{generated_freq_eq}) can also be used for the case of the electron moving with relativistic speed with the only difference that $\omega_L$ must be replaced with $4\gamma^2 \omega_L$, where $\gamma$ is the relativistic gamma factor of the electron.    The $a_0^2$ term in the denominator comes from the fact that an electron is pushed by the electromagnetic wave in the direction of its propagation, and thus moves away from the laser red shifting the reflected light.  Movement of the electron through the focus of the laser and/or pulsing of the laser beam result in $a_0$ being a function of time representing the laser envelope. This in turn means that during the laser pulse interaction with an electron, different frequencies are generated at different times and different electron positions within the envelope, broadening the emitted spectrum. This is illustrated by Fig.~\ref{schematic_figure}~(right).   
 One can also see that certain frequencies are generated twice during the interaction. For example, the frequency $\omega_1$ is generated at two different longitudinal positions of the electron $z_1$ and $z_2$ as shown.  Depending on the value of $\omega_1$ and the separation between the emission points this leads to either constructive or destructive interference  in the generated spectrum. These  interference patterns lead the appearance of  bands in the spectrum.  
 
 The number of oscillations in the spectrum can be approximately established as a ratio of maximum frequency broadening due to laser intensity derived from eq.~(\ref{generated_freq_eq}) and given by $\Delta \omega=\omega_L-\frac{\omega_L}{1+a_0^2}$ and the bandwidth of the laser pulse. Thus, the number of oscillations is roughly given by
 \begin{equation}
 N_{osc}=\omega_L\frac{a_0^2}{1+a_0^2}\frac{1}{\Delta \omega_L}\mathrm{,}
 \end{equation}
 where $\Delta\omega_L$ is the FWHM bandwidth of the laser. One can see that the number of interference fringes in the spectrum grows with increase of both the laser amplitude and laser pulse duration (as laser bandwidth is back proportional to laser duration).  
 
 The exact shape of the spectrum depends on the laser pulse duration and on its envelope shape and intensity,  and can be calculated from the well known motion of a free electron in the plane electromagnetic wave~\cite{Jackson1975,LandauVol2,Sarachik1970}. Here, we neglect the radiation reaction so that the electron dynamics is governed by the standard Lorentz force.  For an electron initially at rest  and for an electromagnetic wave impinging the electron from the $z\rightarrow - \infty$, one can immediately write two integrals of motion for the electron:
\begin{eqnarray}
u_\perp=a_\perp\label{u_perp_eq}\\
\gamma-u_z=1\mathrm{.}
\end{eqnarray}
The latter equation can be also be written in the following form:
\begin{equation}
u_z=\frac{a_\perp^2}{2}\mathrm{.}
\end{equation}
Here $u_\perp=\frac{p_\perp}{m_ec}$ is any perpendicular to $z$-axis electron momentum component, $a_\perp$ - any perpendicular component of the vector potential, $u_z$ is the electron longitudinal momentum component. 
It is easy to show that for the observer looking exactly on-axis, the retarded time that goes into expressions for calculating the spectrum~\cite{Jackson1975,LandauVol2} doesn't depend on $x$ and $y$ coordinates. Hence,  to derive the on-axis spectrum one only needs to know the $z$ coordinate of the electron, which can be found from the following equation:
\begin{equation}
\frac{dz}{d\eta}=u_z
\end{equation}
or, equivalently
\begin{equation}
z(\eta)=\int\limits_{-\infty}^{\eta}u_z(\eta\prime)d\eta\prime=\int\limits_{-\infty}^{\eta}\frac{a_\perp(\eta\prime)^2}{2}d\eta\prime\mathrm{.}\label{z_coordinate_eq}
\end{equation}
Here, $\eta=\omega_L(t-\tilde z)$ is the electron proper time and $z=\frac{\omega_L}{c}\tilde z$ with $t$ and $\tilde z$ being time and longitudinal coordinate in CGS units respectively. 

Since the trajectory of the electron in plane electromagnetic wave is completely known, the on-axis reflected radiation produced by the electron moving in such a wave can be found using the well known formula~\cite{Jackson1975,LandauVol2}:
\begin{equation}
\left.\frac{d^2I}{d\omega d\Omega}\right|_{\theta=0}=\kappa\frac{\omega^2}{4\pi^2}\left|\int\limits_{-\infty}^{+\infty}\mathbf{n}\times\left(\mathbf{n}\times  \mathbf{u_\perp}(\eta)\right)e^{i\omega\left[\eta+2z(\eta) \right]}d\eta\right|^2\mathrm{,}\label{usual_formula_eq}
\end{equation}
where $\Omega$ is the solid angle and the whole formula is evaluated for the on-axis case ($\theta=0$), $\kappa=\frac{e^2\omega_L^2}{c}$ is the normalization coefficient (such that energy spectrum given by eq.~\ref{usual_formula_eq} is given in CGS units) and $ \mathbf{u_\perp}$ is the vector of perpendicular momentum components. 
Using eqns.~(\ref{u_perp_eq}) and (\ref{z_coordinate_eq}) one can rewrite eq.~(\ref{usual_formula_eq}) in terms of the laser amplitude:
\begin{equation}
\left.\frac{d^2I}{d\omega d\Omega}\right|_{\theta=0}=\kappa\frac{\omega^2}{4\pi^2}\left|\int\limits_{-\infty}^{+\infty}\mathbf{n}\times\left(\mathbf{n}\times  \mathbf{a_\perp}(\eta)\right)e^{i\omega\left[\eta+\int\limits_{-\infty}^{\eta}a_\perp^2(\eta\prime)d\eta\prime \right]}d\eta\right|^2\mathrm{.}\label{usual_formula_2_eq}
\end{equation}
For a circularly polarized laser pulse with frequency $\omega_0$ the laser amplitude function can then be expressed as:
\begin{equation}
\mathbf{a_\perp}(\eta)=\frac{1}{2}a(\eta)\cdot \boldsymbol\varepsilon \cdot e^{i\phi(\eta)}+\mathrm{c.c.,}\label{total_pulse_eq}
\end{equation}
where $a(\eta)$ is the envelope function of the pulse and $\boldsymbol\varepsilon=\mathbf{e_x}+i\mathbf{e_y}$ is introduced to take into account circular polarization and $\phi(\eta)=\omega_0(\eta)\eta$. We will call $\omega_0=\phi(\eta)/\eta$ - frequency of the laser pulse and $\omega_i(\eta)=\frac{d\phi}{d\eta}$ - instantaneous frequency of the pulse.  In the case of a laser pulse with constant frequency $\omega_0=1$ due to our choice of units, but we keep it in these equations to facilitate consideration of cases with chirped laser pulses.

Expressions for the on-axis spectrum can be obtained
 in the fully nonlinear case
  for a laser having an envelope in time described by a half-sine profile:
\begin{eqnarray}
a(\eta)=a_0\cdot \sin\left[\frac{\pi \eta}{\tau_L}\right]\mathrm{,}\quad\quad0<\eta<\tau_L\mathrm{,}\label{laser_envelope_eq}
\end{eqnarray}
with $\tau_L$ being the duration of the laser pulse.  For such a pulse, the $z$ coordinate of the electron can be analytically found from eq.~(\ref{z_coordinate_eq})
\begin{equation}
z(\eta)=\frac{a_0^2}{4}\left(\eta-\frac{\tau_L}{2\pi}\sin\frac{2\pi\eta}{\tau_L} \right)\mathrm{.}\label{z_eq}
\end{equation}
The on-axis spectrum of reflected radiation can then be rewritten in the following form
\begin{equation}
\left.\frac{d^2I}{d\omega d\Omega}\right|_{\theta=0}=\kappa\omega^2\frac{a_0^2}{2\omega_0^2}N_0^2\left|\int\limits_0^{2\pi} \sin\left(\frac{\xi}{2}\right)\sin\left(N_0\xi \right)e^{i\tilde\omega N_0\xi-i \chi \sin\xi }d\xi\right|^2\mathrm{,}
\label{integral_eq}
\end{equation}
where
\begin{eqnarray}
N_0=\frac{\omega_0\tau_L}{2\pi}\mathrm{,}\quad\chi=\frac{\omega a_0^2\tau_L}{4\pi}\mathrm{,}\quad \tilde\omega=\frac{\omega}{\omega_0}\left(1+\frac{a_0^2}{2} \right)\mathrm{.}
\end{eqnarray} 

\begin{figure}[h!!]
\subfigure{\includegraphics[width=0.4\textwidth]{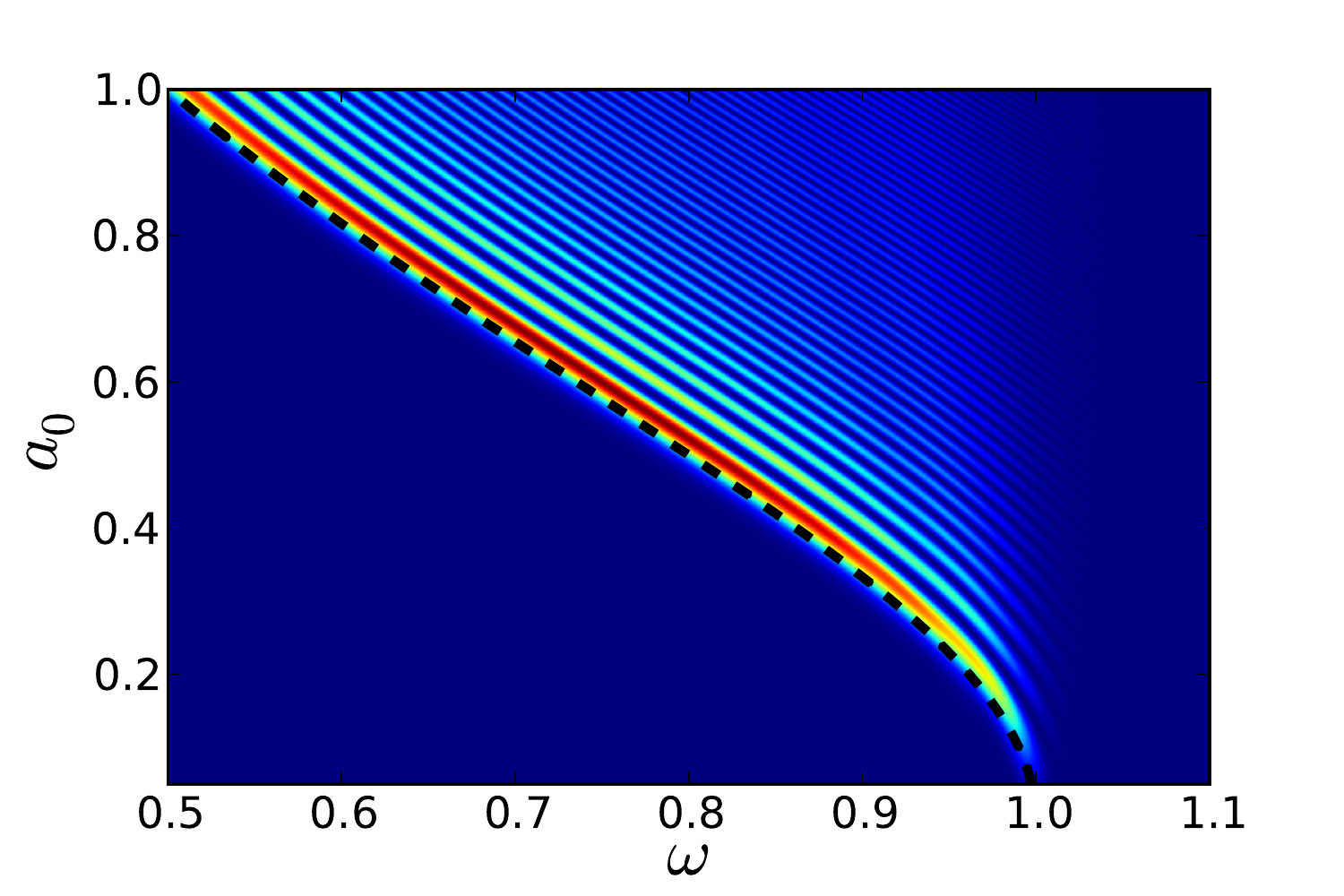}}
\subfigure{\includegraphics[width=0.4\textwidth]{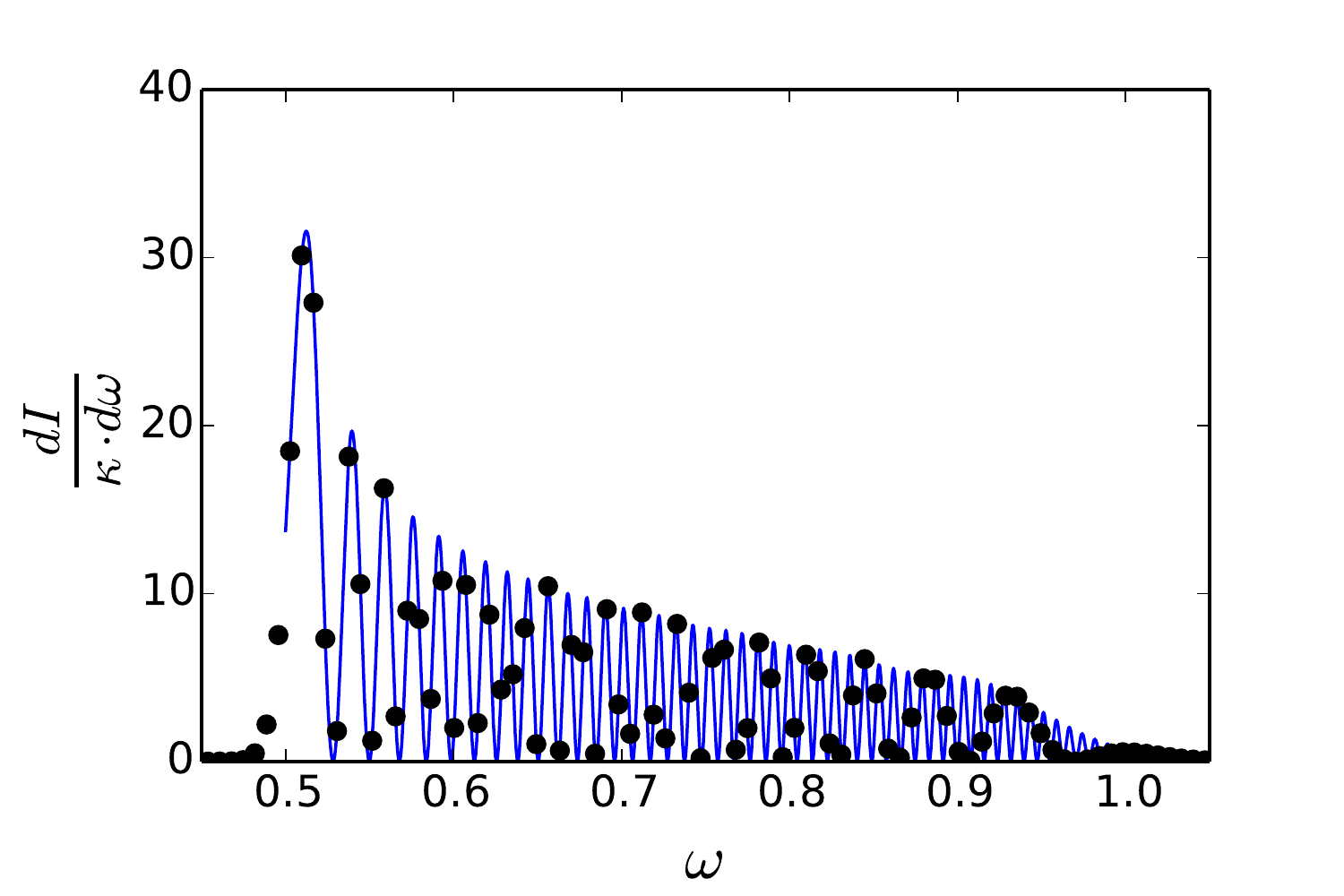}}
\caption{Left: On-axis radiation spectra plotted as a function of laser pulse amplitude $a_0$ (vertical axis).  The spectra are in arbitrary units, from numerical integrationof eq.~(\ref{integral_eq}). Laser pulse  duration is $\tau_L=600$.  The dashed black line represents the red-shift of the peak of the spectrum according to eq.~(\ref{generated_freq_eq}).  Right: On-axis radiation spectrum for $a_0=1$ from the analytical expression of eq.~(\ref{spec_bessel_an_eq})~(black circles) for such frequencies as make the order of the Bessel function given by eq.~(\ref{bessel_order_eq}) an integer, overplotted with the full spectrum from numerical integration   (blue line).}
\label{numerical_an_spec_figure}
\end{figure}

Numerical integration of these equations can be used to show the dependence of the spectrum on $a_0$, as illustrated by a color-coded image of the on-axis spectrum in Fig.~\ref{numerical_an_spec_figure}(left).  The spectra are obtained from numerical integration of eq.~(\ref{integral_eq}) for different $a_0$ (vertical axis) and for laser pulse with duration $\tau_L=600$.  One can see that for low values of $a_0$ the spectrum is narrow and is limited by the bandwidth of the incoming electromagnetic wave, whereas for large values of $a_0$ the spectrum is broad and band sub-structure is visible. The main peak of the spectrum is red-shifted and its position is given by eq.~(\ref{generated_freq_eq}) (shown with dashed line on the Fig.~\ref{numerical_an_spec_figure}~(left)). 

While numerical integration of the nonlinear spectrum has been conducted previously, analytic solutions are important to allow understanding of the mechanisms of broadening and to allow design of techniques to compensate for broadening.  Opening the sine functions using the Euler's formula one can analytically evaluate the integral to obtain directly the nonlinear spectrum.  Doing so yields:
\begin{equation}
\left.\frac{d^2I}{d\omega d\Omega}\right|_{\theta=0}=\kappa\left(\frac{\omega}{\omega_0}\right)^2\frac{a_0^2}{8}N_0^2\cdot\left|J_n(\chi)-J_{n-1}(\chi) \right|^2\mathrm{,}\label{spec_bessel_an_eq}
\end{equation}
where $J_n$ is the Bessel function and
\begin{equation}
 n=N_0(\tilde\omega-1)+\frac{1}{2}\mathrm{.}\label{bessel_order_eq}
 \end{equation}
 Here we have neglected two small terms that are given by Bessel functions with orders $N_0(\tilde\omega+1)\pm\frac{1}{2}$  because $N_0\gg 1$ typically for $\tau_L\gg 2\pi/\omega_0$ and Bessel function with such orders is very small in the frequency range of interest. 
 It is important to note that the solution is valid only for a subset of all $\omega$ where the order $n$ is integer. For other values of $\omega$, numerical integration of eq.~(\ref{integral_eq}) is required. Fig.~\ref{numerical_an_spec_figure}(right) displays the spectrum for $a_0=1$ and laser pulse with duration $\tau_L=600$ . Black dots show the analytical solution from eq.~(\ref{spec_bessel_an_eq}) for the frequencies $\omega$ that make the order of the Bessel function an integer. One can see that analytical solution fits the numerical integration  (blue line)  very well.  Although the analytical solution provides only a discrete set of points, it well  outlines the shape of the spectrum and gives its peak value and width which are the most important parameters for many applications.  This is especially true because in many Thomson scattering photon source applications the fine-scale oscillations observable for a single electron will be washed out by the nonzero energy spread of the electron beam.  In the example shown here, the oscillations have   frequency spacing  of $\simeq$2$\% \omega_0$ which will be washed out by electron energy spread of $\simeq$1$\%$.
  The analytic expressions then allow us to analytically evaluate broadening, and give us a tool to calculate compensating terms which can be used to control and narrow the spectrum. 

\section{Controlling the shape of the spectrum by laser pulse chirping}\label{chirping_sec}

Nonlinear broadening of the spectrum is a result of red-shifting of the Thomson scattered frequency during the interaction due to the longitudinal motion of the electron, as given by eq.~(\ref{generated_freq_eq}) with $a_0$ being a function of time representing the laser pulse envelope. This indicates that by compensating the laser frequency by chirping the laser pulse, one can diminish  broadening of the generated spectrum. This was proposed, and simulation based on a single set of numerical parameters was provided in~\cite{Ghebregziabher2013}, and further examples were provided by ~\cite{Terzic2014}.  Here we extend the analytical expressions obtained above to derive analytical expressions for the radiation spectrum in the case of proper laser pulse chirping.  Using appropriate laser chirp, spectrum narrowing can be obtained for arbitrary values of laser pulse amplitude $a_0$. The analytical results are compared with numerical integration.

We consider again the circularly polarized laser pulse with the envelope given by eq.~(\ref{laser_envelope_eq}). If the frequency of the laser pulse is constant the generated frequency is given by eq.~(\ref{generated_freq_eq}), and the resulting spectra are those of Fig.~\ref{numerical_an_spec_figure}. However, if the instantaneous frequency of the laser pulse is given by
\begin{equation}
\omega_{i}(\eta)=\left[1+a(\eta)^2 \right]\mathrm{,}\label{chirp_eq}
\end{equation}
one can expect that the generated frequency will be constant and equal to one independent of the value of $a_0$. This can be derived from the following considerations. Substituting eq.~(\ref{total_pulse_eq}) into eq.~(\ref{usual_formula_2_eq}) and looking at the oscillatory term of the integral $e^{i\Phi(\eta)}$, one can write for the $\Phi(\eta)$
\begin{equation}
\Phi(\eta)=-\omega_0(\eta)\eta+\omega\eta+\omega\int_{-\infty}^{\eta}a_\perp^2(\eta\prime)d\eta\prime\mathrm{.}
\end{equation}
We are looking for such frequency function $\omega_0(\eta)$ that will compensate the red-shifting and force the maximum of the integral to be at $\omega=1$. In order to do that $\Phi(\eta)$ for $\omega=1$ must be constant, i.e. $C$, so that the integral has the maximum value. Thus,
\begin{equation}
-\omega_0(\eta)\eta+\eta+\int_{-\infty}^{\eta}a_\perp^2(\eta\prime)d\eta\prime=C
\end{equation}
Writing $\omega_0(\eta)=1+f(\eta)$, one can find that $f(\eta)\eta=\int_{-\infty}^{\eta}a_\perp^2(\eta\prime)d\eta\prime-C$. For the phase of the properly chirped laser pulse one can thus write
\begin{equation}
\phi(\eta)=\eta+\int_{-\infty}^{\eta}a_\perp^2(\eta\prime)d\eta\prime-C\mathrm{,}\label{laser_phase_eq}
\end{equation}
and for the instantaneous laser pulse frequency one obtains the following expression
\begin{equation}
\omega_i(\eta)=\frac{d\phi(\eta)}{d\eta}=1+a_\perp^2(\eta)\mathrm{.}\label{instantaneous_freq_eq}
\end{equation}
For the laser pulse with envelope given by eq.~(\ref{laser_envelope_eq}) and instantaneous frequency given by eq.~(\ref{instantaneous_freq_eq}) one can get analytical solution.
Note, that the longitudinal coordinate of the electron is still given by eq.~(\ref{z_eq}) as the laser pulse is chosen to be circular.
The result of the spectrum calculation using eq.~(\ref{usual_formula_eq}) yields the same formula as in eq.~(\ref{spec_bessel_an_eq}), but with $n$ and $\chi$ given by
\begin{eqnarray}
n=\left(\omega-1 \right)N_0\cdot\left(1+\frac{a_0^2}{2} \right)+\frac{1}{2}\label{n_chirp_eq}\\
\chi=\frac{1-\omega}{2}N_0\label{chi_chirp_eq}\mathrm{.}
\end{eqnarray}

\begin{figure}[h!!]
\subfigure{\includegraphics[width=0.4\textwidth]{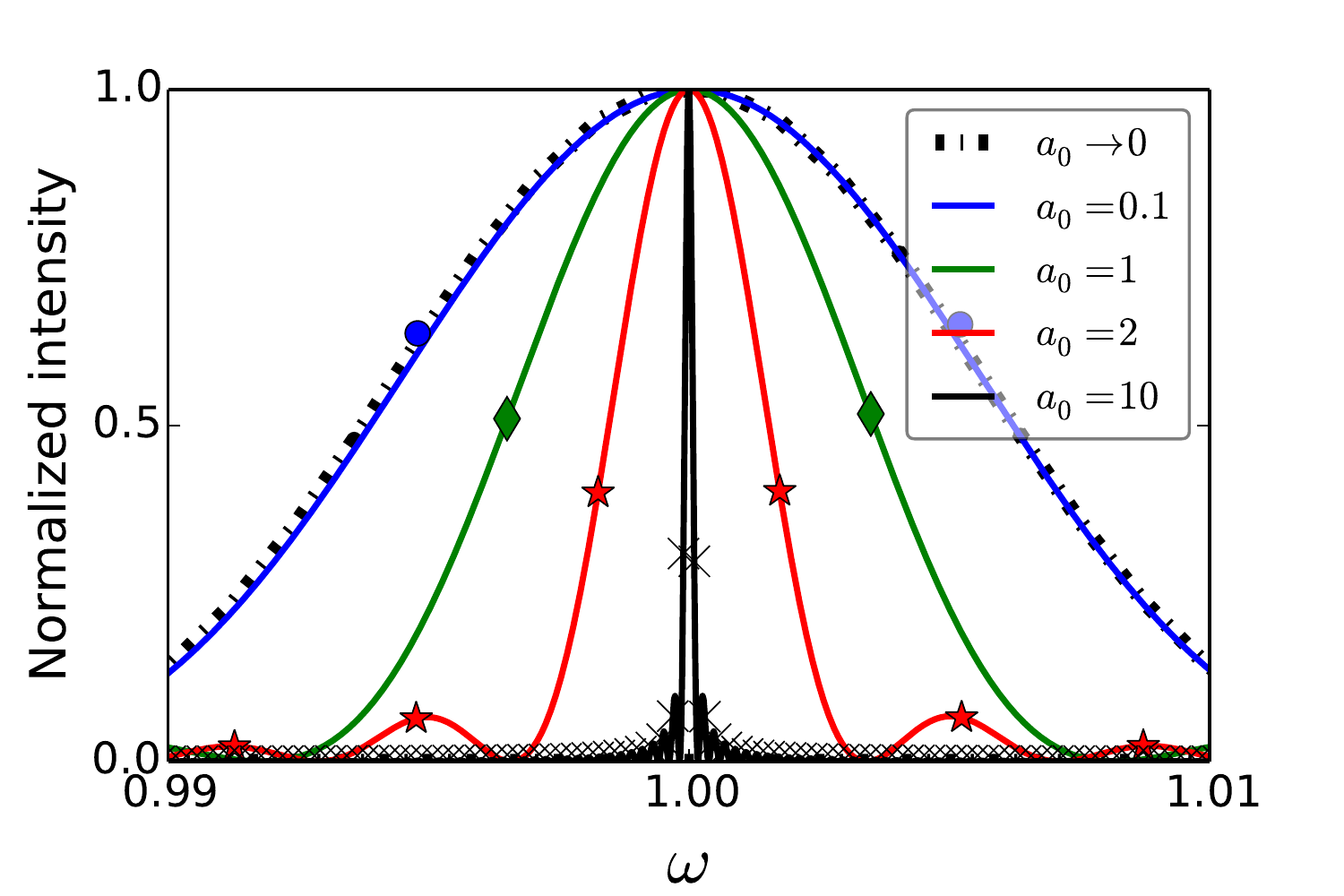}}
\subfigure{\includegraphics[width=0.4\textwidth]{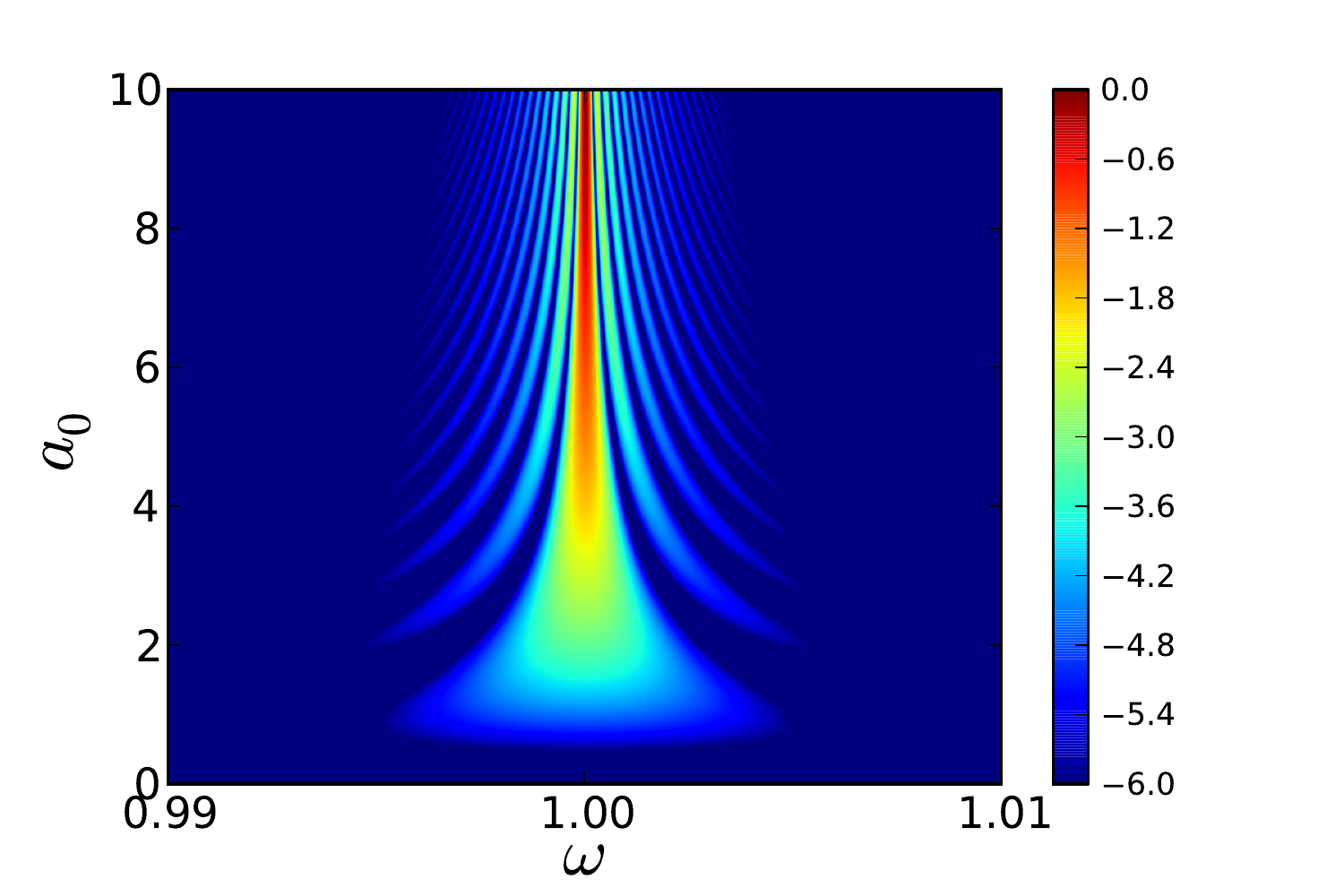}}
\caption{Left: On-axis radiation spectra for the laser pulse that is properly chirped according to eq.~(\ref{chirp_eq}), and with other   parameters as in Fig.~\ref{numerical_an_spec_figure}. The results are plotted in the logarithmic scale. Right: On-axis spectra for different values of $a_0$ obtained from numerical integration (solid lines) and analytical expression (markers of corresponding color) using eqns.~(\ref{spec_bessel_an_eq}), (\ref{n_chirp_eq}) and (\ref{chi_chirp_eq}).}
\label{numerical_an_spec_chirp_figure}
\end{figure}

\begin{figure}[h!!]
\subfigure{\includegraphics[width=0.4\textwidth]{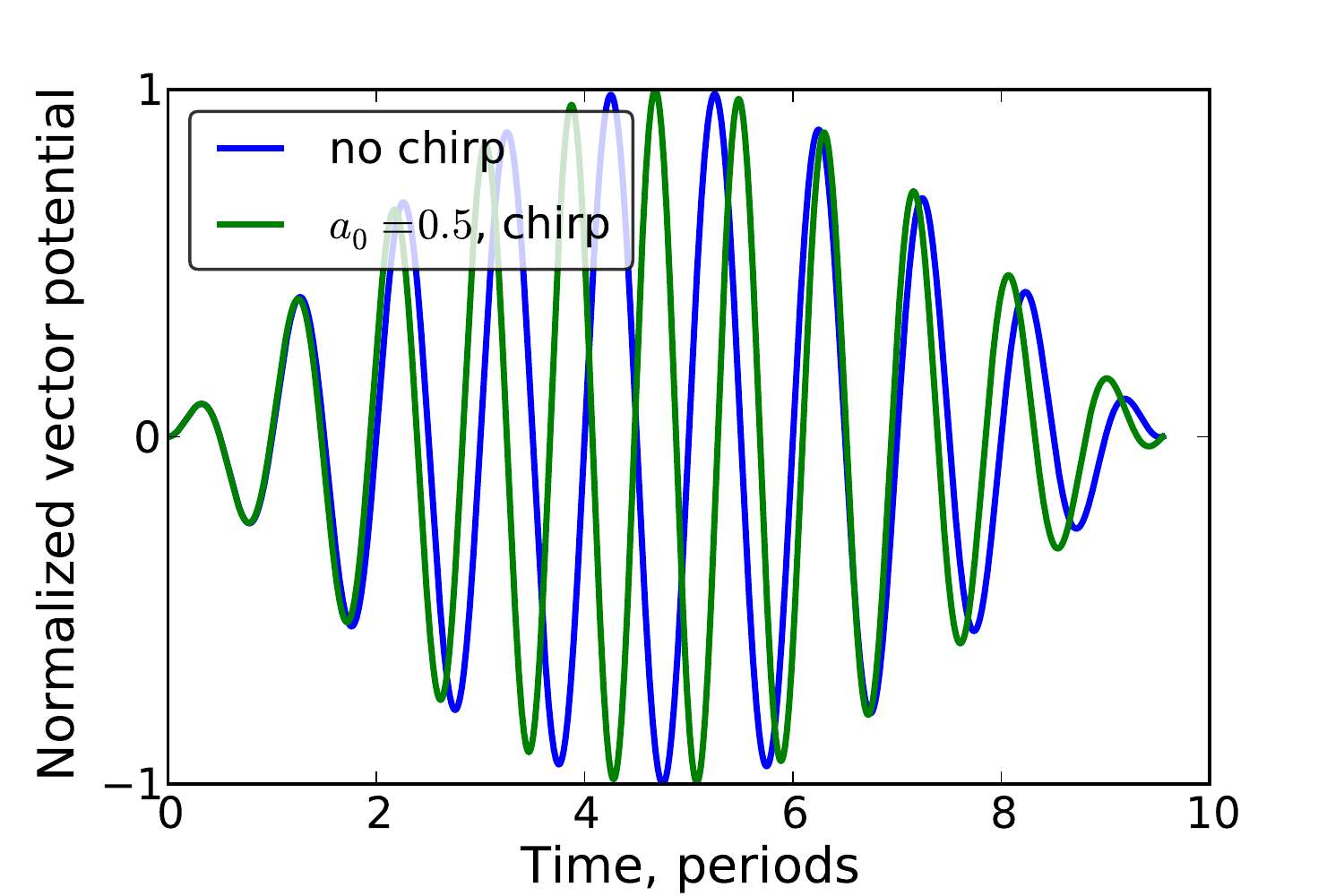}}
\subfigure{\includegraphics[width=0.4\textwidth]{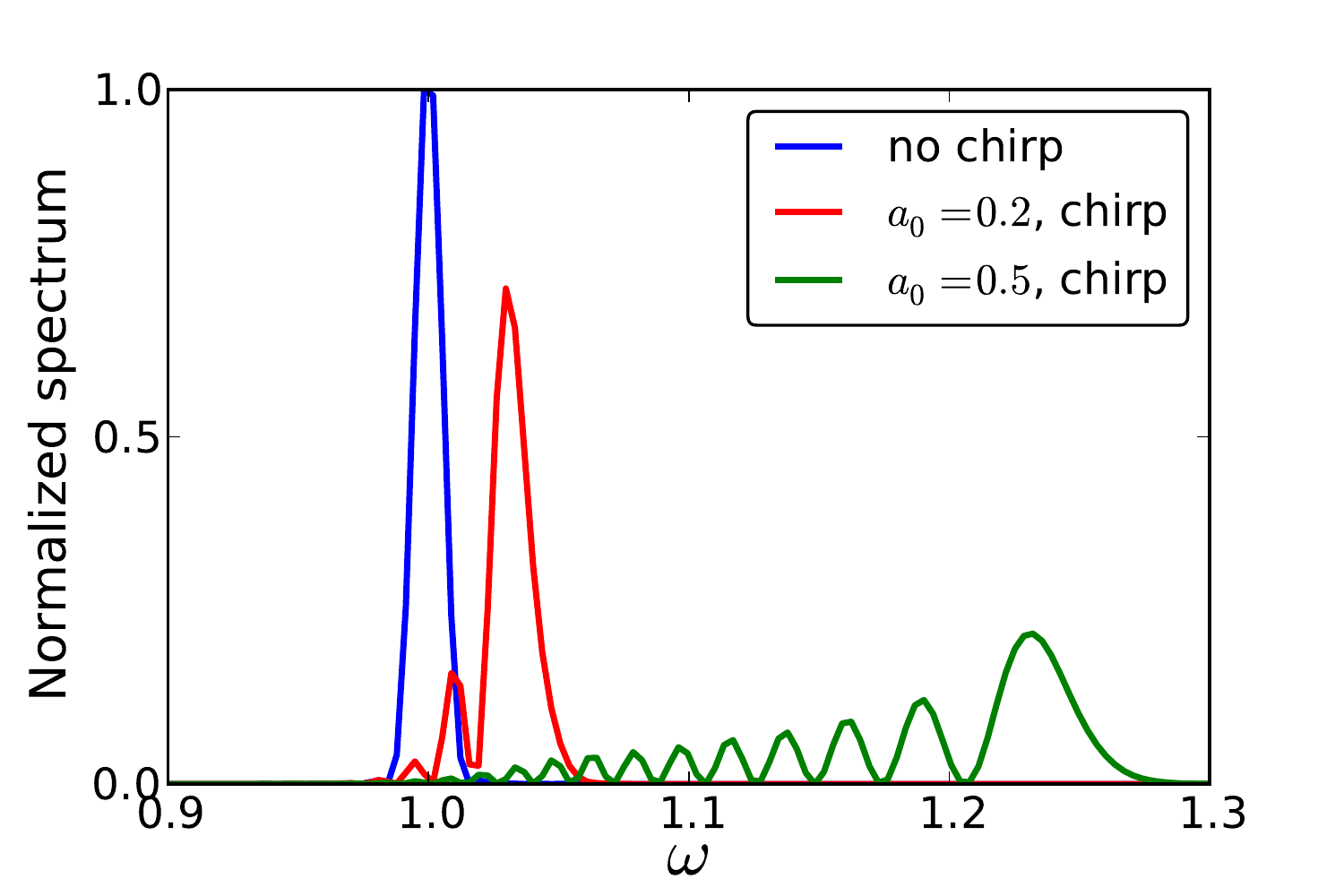}}
\caption{Left: Normalized vector potential as a function of time in periods for the case of the unchirped laser pulse (blue color) and chirped laser pulse with $a_0=0.5$ for laser pulse with duration $\tau_L=60$. Right: Laser pulse spectra (normalized to the peak value of the unchirped pulse spectrum) for the case of unchirped laser pulse (blue color) and chirped laser pulses with $a_0=0.2$ (red color) and $a_0=0.5$ (green color) for laser pulse with duration $\tau_L=600$.}
\label{vector_potential_figure}
\end{figure}

 The analytical solutions for the nonlinear spectral bandwidth using  a laser pulse with the envelope given by eq.~(\ref{laser_envelope_eq}) and instantaneous frequency changing under the law of eq.~(\ref{chirp_eq}) are shown in Fig.~{\ref{numerical_an_spec_chirp_figure}}~(left), where normalized spectra are presented for different values of $a_0$. Markers of different colors  show the analytical solutions using eqns.~(\ref{spec_bessel_an_eq}), (\ref{n_chirp_eq}) and (\ref{chi_chirp_eq}), while solid lines of corresponding colors show the numerical integration results. In the case of $a_0=0.1$ (blue color) the spectrum width is approximately given by the unchirped (i.e. with constant frequency, shown on the figure with black dash-dot line) laser pulse width, which is back proportional to laser pulse duration $\tau_L$.
  In the case of $a_0=10$ (black color) the spectrum width is approximately $a_0^2=100$ times narrower. Analytical solutions fit with numerical integration well and predict spectrum narrowing for properly chirped laser pulses with arbitrarily high $a_0$.  The generated frequency stays centered at $\omega=1$ and the broadening disappears. One can see that the spectrum is actually getting narrower with the increase of $a_0$. This is due to the choice of the laser pulse function. Indeed, changing (increasing compared to $\omega=1$) the frequency while keeping the duration $\tau_L$ constant leads to more sign changes of the vector potential of the laser pulse or, in other words, to more periods. This can be seen from Fig.~\ref{vector_potential_figure}~(left) where normalized laser pulse vector potential is plotted for the case when the laser pulse is unchirped (blue color) and for the case of chirped laser pulse with $a_0=0.5$ (green color). In this figure the duration of the laser pulse was set to $\tau_L=60$ for  illustrative reasons. Because at each period an electron is forced (by appropriate chirp) to radiate the same frequency, it effectively radiates pulses with the same frequency $\omega=1$ but with longer duration for higher $a_0$ leading to narrower spectrum.  Numerical integration of the generated spectra using eq.~(\ref{usual_formula_eq})  is presented in Fig.~\ref{numerical_an_spec_chirp_figure} (right).  The color-coded image is the normalized on-axis spectrum (in logarithmic scale) as a function of both the frequency (longitudinal axis) and normalized laser amplitude $a_0$ (vertical axis) similar to Fig.~\ref{numerical_an_spec_figure}~(left).

It is worth noting that the spectrum of the incident chirped pulse extends approximately up to the frequency $\omega_{max}=\left(1+a_0^2 \right)$,  quadratically with $a_0$. Figure~\ref{vector_potential_figure}~(right) shows the normalized spectra of laser pulses properly chirped according to eq.~(\ref{chirp_eq}) (red color corresponds to $a_0=0.2$ and green color corresponds to $a_0=0.5$) compared to the case of the unchirped laser pulse.  For low values of $a_0<1$ the introduced chirp can be on the order of 10-20 percent and is achievable with current technology.   This already allows production of narrow bandwidth sources using significantly higher laser intensity than is conventionally possible, which in turn reduces the required laser energy.  As noted above, even operation at $a_0$=0.3 can save a factor of four in scattering laser energy.     While in principle the technique can be used up to even higher intensities, practical implementation is limited by the obtainable bandwidth in the scattering laser.  For example, at  $a_0=10$ the laser pulse contains the range of wavelengths from x-rays to the laser wavelength, which is beyond currently foreseeable  laser technology. 

\section{Conclusions}\label{discussion_sec}
In this paper we have presented analytical solutions for the spectrum of radiation generated by a free electron interacting with a circularly polarized laser pulse of nonlinear intensity.  The results are derived for  an electron initially at rest for computational clarity.  They apply also, using a straightforward Lorentz transform into the beam frame, to Thomson scattering sources of x-rays and gamma rays which scatter laser pulses from relativistic electron beams.  The analytical results agree very well with results of numerical integration and  provide useful insights and scalings for  nonlinear Thomson scattering.   We have shown analytically and numerically that by proper chirping of the laser pulse the broadening of the radiation spectrum can be avoided for arbitrarily laser laser pulse amplitude $a_0$. The results predict successfully laser amplitude and pulse shape parameters to compensate nonlinear broadening and produce narrow bandwidth sources at high intensity.  This result is important for generation of high flux Thomson scattering sources of x-rays and gamma rays, as it allows generation of a given photon flux using greatly reduced laser energy.  Moreover, the results presented in this paper can be used for optimization of experiments as well as benchmarking of the numerical tools.

\section*{Acknowledgements}
This work was supported by the U.S. Dept. of Energy National Nuclear Security administration DNN R\&D/NA-22, and by the Office of Science Office of High Energy Physics,  under Contract No. DE-AC02-05CH11231. The simulations used the resources of the National Energy Research Scientific Computing Center, a DOE Office of Science User Facility supported by the Office of Science of the U.S. Department of Energy under Contract No. DE-AC02-05CH11231.  We would like to acknowledge fruitful discussions with M.~Zolotorev, M.~Efremov and C.~Benedetti.







\end{document}